\documentclass[sigconf]{acmart}
\usepackage{balance}

\AtBeginDocument{%
  \providecommand\BibTeX{{%
    \normalfont B\kern-0.5em{\scshape i\kern-0.25em b}\kern-0.8em\TeX}}}

\copyrightyear{2020} 
\acmYear{2020} 
\setcopyright{acmlicensed}
\acmConference[SIGCSE '20]{The 51st ACM Technical Symposium on Computer Science Education}{March 11--14, 2020}{Portland, OR, USA}
\acmBooktitle{The 51st ACM Technical Symposium on Computer Science Education (SIGCSE '20), March 11--14, 2020, Portland, OR, USA}
\acmPrice{15.00}
\acmDOI{10.1145/3328778.3366907}
\acmISBN{978-1-4503-6793-6/20/03}

\begin{document}
	
\fancyhead{}

\title{Securing Bring-Your-Own-Device (BYOD) Programming Exams}

\author{Oka Kurniawan}
\orcid{0000-0001-9519-0959}
\affiliation{\institution{Singapore University of Technology and Design}}
\email{oka_kurniawan@sutd.edu.sg}

\author{Norman Tiong Seng Lee}
\orcid{0000-0003-4288-3588}
\affiliation{\institution{Singapore University of Technology and Design}}
\email{norman_lee@sutd.edu.sg}

\author{Christopher M. Poskitt}
\orcid{0000-0002-9376-2471}
\authornote{Work done while affiliated with the Singapore University of Technology and Design.}
\affiliation{\institution{Singapore University of Technology and Design}}
\affiliation{\institution{Singapore Management University}}

\newcommand{\substepseparator}{\hspace{1cm}}

\begin{abstract}
	Traditional pen and paper exams are inadequate for modern university programming courses as they are misaligned with pedagogies and learning objectives that target practical coding ability. Unfortunately, many institutions lack the resources or space to be able to run assessments in dedicated computer labs. This has motivated the development of bring-your-own-device~(BYOD) exam formats, allowing students to program in a similar environment to how they learnt, but presenting instructors with significant additional challenges in preventing plagiarism and cheating. In this paper, we describe a BYOD exam solution based on \emph{lockdown browsers}, software which temporarily turns students' laptops into secure workstations with limited system or internet access. We combine the use of this technology with a learning management system and cloud-based programming tool to facilitate conceptual and practical programming questions that can be tackled in an interactive but controlled environment. We reflect on our experience of implementing this solution for a major undergraduate programming course, highlighting our principal lesson that policies and support mechanisms are as important to consider as the technology itself.
\end{abstract}

\ccsdesc[500]{Applied computing~Education}
\ccsdesc[500]{Social and professional topics~Student assessment}

\keywords{Programming exams; BYOD exams; lockdown browsers; learning management systems; cloud-based IDEs; plagiarism prevention}

\maketitle

\section{Introduction}
University courses have traditionally been assessed by written examinations: pen and paper at the ready, separated desks, a clock counting down, and invigilators pacing the room~\cite{Sheard-et_al13a}. This format has survived the test of time because it is simple for instructors to administer, has well-established logistics, and perhaps most importantly, is run in a highly controlled environment, minimising the risk and temptation of cheating and plagiarism.

In a modern computer science curriculum, however, this style of assessment is completely misaligned with pedagogies and learning objectives that target \emph{practical programming ability}. In an introductory programming course, for example, students learn by interacting with a language's compiler or interpreter: trial and error, testing, debugging, and even looking things up in documentation are all part of the programming experience, regardless of ability. Yet a traditional written exam for such a course is limited to testing the concepts, or the ability to `code' on pen and paper, forcing instructors to simplify the questions and forcing students to train for the exam. While adding a project component to the course can alleviate this problem, retaining some kind of final exam remains a popular option for assessing individual learning outcomes of students.

Ideally, the exam of a programming course should recreate the environment that students learnt and practiced in, e.g.~by providing access to their Integrated Development Environment~(IDE) of choice. One way to achieve this is to run the exam in a dedicated computer laboratory in which user accounts have reduced privileges and are unable to connect to the internet. This solution has been demonstrated as effective (e.g.~\cite{English02a,Daly-Waldron04a}), and Zilles et al.~\cite{Zilles-et_al18a} have shown that it is possible to run lab-based testing centres at a reasonable cost. Many institutions, however, lack the resources or space to be able to implement such facilities at the scale required. An alternative option is to run a \emph{bring-your-own-device~(BYOD)} exam in which students use their own laptops to complete the programming questions (e.g.~\cite{Hillier-Fluck17a,Stephenson18a}). This format allows for students to complete the exam using IDEs and tools they are familiar with. However, without the right technology and policies in place, it is difficult for invigilators to prevent plagiarism and cheating~\cite{Sindre-Vegendla15a}. 

In this article, we describe a BYOD exam solution based on \emph{lockdown browsers}, software which temporarily turns students' laptops into secure workstations with limited access to applications, system functions, or the internet. Our solution combines this technology with a Learning Management System~(LMS) and web-based IDE to facilitate conceptual and practical programming questions in an interactive but controlled environment. We describe how we implemented it for a major first-year undergraduate programming course (approx.~500 students), and reflect on the experiences and comments of both the instructors and students. Overall, we found our BYOD solution to be a reasonable compromise between our goal of aligning programming assessments with course pedagogies and our goal of ensuring the security and integrity of the exam. Furthermore, while technology is clearly key to our solution, we learnt that policies and support mechanisms are equally important considerations to ensure the success of BYOD programming exams.

This paper is a revised and extended version of a short article featured in the pedagogy newsletter of the authors' institution~\cite{Kurniawan-Lee-Poskitt19a}.



\section{Related Work}\label{sec:related_work}

In general, instructors are aware of the need to align their assessments, learning objectives, and pedagogies~\cite{Biggs96a}. Many studies report that practical exams for programming help to achieve this and a number of other benefits. First, they allow students to demonstrate their programming skills in a setting that is close to how they typically code day-to-day~\cite{Carrasquel85a, Bennedsen-Caspersen07a}. Second, the grading of student answers can often be carried out by automated assessment tools, enabling feedback to be provided in a shorter time compared to manual marking~\cite{Daly-Waldron04a,Stephenson18a}. Third, students are reported to be positive about having access to a compiler, as it enables their code to be tested and syntax errors to be caught~\cite{Stephenson18a,Zilles-et_al15a}. Finally, having practical exams allows data on students' answers to be collected for further analysis of errors that students make~\cite{English02a}.

Several studies have described the implementation of practical end-of-course exams for programming, including two that did so using computer laboratories in a single sitting. Stephenson~\cite{Stephenson18a} describes how a programming exam was administered for 109 students, each of whom were allowed full internet access, but with communication between them deterred by having the invigilators provide close supervision. Students were given the option of using their own laptop or the lab computers, with most students opting for the latter. Exam answers that did not pass all the test cases in the autograder were manually graded. Daly and Waldron~\cite{Daly-Waldron04a} reported a programming exam administered to almost 400 students simultaneously using a dedicated computer laboratory. Automated grading was also used. In their implementation, special student accounts that blocked internet access and had reduced privileges were used to prevent plagiarism (an approach also taken by English~\cite{English02a}). Students were, however, allowed to bring along a pre-prepared `cheat sheet' to the laboratory. 

Another approach described by some studies is to conduct such programming exams during regular timetabled lab sessions, thus requiring multiple sittings. Califf and Goodwin~\cite{Califf02a} conducted their final exam in 14 sittings over four days, requiring several versions of the exam question to be written. Instead of a final exam, other studies describe assessing students using regular lab exams scheduled over the duration of the course~\cite{Jacobson00a, Barros03a}. There can be many variations in the conduct of laboratory-based programming exams. Cutts et al.~\cite{Cutts-et_al06a}, for example, describe how such exams were run in seven different UK universities.

Dedicated computer laboratories require significant investment, and are not practical for exams with large numbers of students. Running exams on students' own laptops, commonly termed as `bring-your-own-device electronic exams' or `BYOD (e-)exams', is another solution. Hillier and Fluck~\cite{Hillier-Fluck17a} describe how a secure and standard operating system environment can be provided by giving each student a Linux USB stick to boot up their computer. Ribeiro and Amaral~\cite{Ribeiro-Amaral18a} required students to install the open source Safe Exam Browser~\cite{SafeExamBrowser} software to restrict network access in their multiple-choice exam. Seow and Soong~\cite{Seow-Soong14a} also administered a multiple-choice BYOD exam, but over two sittings for more than 600 students, and with a prior mock test in which students were able to download the lockdown browser and familiarise themselves with the procedures. Note that neither of these exam formats involved practical programming. ExamSoft~\cite{ExamSoft} and Respondus LockDown Browser~\cite{Respondus} are alternative and commercially available lockdown browsers. Dawson~\cite{Dawson16a} points out some ways to hack such systems and suggests some mitigation strategies, some of which include not distributing exam papers on USB devices, making examinations open book, and making past-year exam papers available to students.

Rajala et al.~\cite{Rajala-et_al16a} developed an in-house examination platform and used it to conduct practical programming exams in Java, with around 100 students at each sitting. On top of a feature to compile and run Java code, the platform also allowed other types of questions to be asked in an exam, such as multiple-choice ones or Parson's Puzzles~\cite{Parsons-Haden06a}. Internet access was restricted via a firewall. It was reported that the exam was ``done supervised in a lecture hall or computer lab'', thus it is unclear whether exams were conducted solely on a BYOD basis or whether lab computers were used too.

\section{Context}\label{sec:context}
The motivation for designing a BYOD solution came directly from ``Digital World''~\cite{DigitalWorld}, our institution's course on computational thinking and programming using Python. In the following, we present an overview of the course, how the students are assessed, and the limitations of its previous exam formats that we set out to overcome.

Digital World is a compulsory introductory programming course taken by \emph{every} first-year undergraduate at the Singapore University of Technology and Design, regardless of the subject they later choose to major in. The course uses elements of the flipped classroom paradigm, requiring students to complete a set of short readings and a corresponding quiz on our LMS before the first class of each week. During class hours, students are given a weekly problem set, containing programming problems that instructors may use as the basis of their lessons, and homework problems that are left to the students to tackle on their own or with their peers. Solutions to these problems are submitted to Vocareum~\cite{Vocareum}, which automatically grades students' solutions according to a variety of test cases. As well as a submission platform, Vocareum is a web-based IDE that students can use to write, run, and debug Python programs.

During the term, students are assessed at multiple junctures via quizzes, homework, and group projects. Nonetheless, a significant portion (50\%) of their final grade is derived from their performance in a mid-term and final exam. Both of these exams are split into two parts: Part~A consists of short, traditional questions that aim to test their understanding of programming \emph{concepts}, whereas Part~B uses practical programming exercises to test whether students can put those concepts into \emph{practice}. Given the large number of students (approx.~500) and limited lab space at our institution, our online exams required a BYOD solution from the outset, which has evolved significantly over the many iterations of the course.

In the mid-term exam of the very first run of the course, students were allowed completely unrestricted use of the internet---similar to Stephenson~\cite{Stephenson18a}. This simplified the setup of the exam, with students able to use their own devices and programming environments without any additional steps. However, despite the presence of invigilators, this unrestricted format unfortunately facilitated some cheating, with some students using file sharing platforms to send solutions to each other. This was detected after the exam when a plagiarism detector was run on the submission platform.

In subsequent exams, it was decided that internet access would be restricted to websites on a whitelist, such as the submission platform itself. Unfortunately, this still presented two major problems. First, for technical reasons, students who were already connected to the router \emph{before} whitelisting was applied would still retain unrestricted access afterwards. Students were required by policy to reset their connections or laptops prior to exams, but there was no straightforward way of checking whether students actually complied with this. Second, it was still possible for students to circumvent our restrictions by connecting to mobile hotspots with hidden SSIDs. This was very difficult for invigilators to check in an exam setting without being highly intrusive. A better BYOD format was required that balanced the benefits with the need for security.

\section{Solution}\label{sec:solution}

\begin{figure}[!t]
	\centering
	\includegraphics[width=0.75\linewidth]{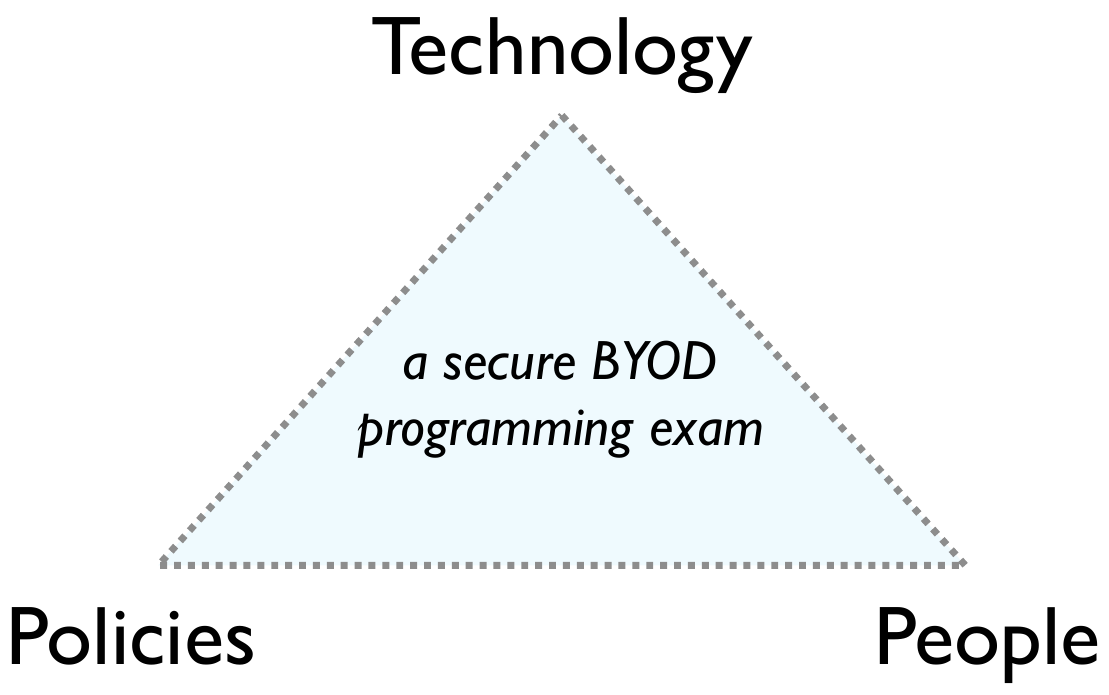}
	\caption{Components of a secure BYOD programming exam}
	\label{fig:aspects}
\end{figure}

In designing a secure BYOD programming exam solution, we identified three interdependent components (Figure~\ref{fig:aspects}) that needed to be addressed. What \emph{technology} do we need to secure our students' devices? What \emph{policies} are needed to safeguard exams? And finally, how will we support the \emph{people} running the exams?

In the 2019 iteration of Digital World, we designed and implemented a comprehensive BYOD solution that addressed all three components. In terms of technology, we introduced the use of a lockdown browser to secure students' laptops while providing controlled access to our LMS, Vocareum, and some additional web-based IDEs. Complementing this technology, we developed policies to ensure a consistent standard of invigilation across examination rooms, and systematic procedures to be followed for different kinds of system failures. Finally, we designed briefing sessions for our invigilators, and benefited from the real-time technical support of both an in-house team (for the lockdown browser and LMS) and remote engineers (for Vocareum).

\substepseparator

\noindent\textbf{Technology.} Our technological solution makes use of Safe Exam Browser~(SEB)~\cite{SafeExamBrowser}, a lockdown browser that temporarily turns students' laptops into secure workstations, limiting their access to unauthorised websites, applications, or system functions. For our exams, we configure SEB to block access to all sites other than those on a whitelist, which consists only of our LMS, Vocareum, and some additional web-based IDEs provided as backup options for students to program in. In our LMS (Blackboard Learn), students answer conceptual questions (Part A), typically structured in formats supporting automated grading, e.g.~multiple choice, multiple answers, matching, ordering, fill in the blank, true/false. Testing specific programming concepts (and not just problem solving) is an approach recommended by Zingaro et al.~\cite{Zingaro-et_al12a}. With Vocareum~\cite{Vocareum}, students interactively develop, test, and submit solutions to practical programming exercises (Part B) in a controlled environment. The platform is familiar to students as it is also used for submitting solutions to problem sheets from our weekly classes. Furthermore, as for these classes, we configure the Vocareum workspace with several executable test cases for each question so that students can quickly check the progress of their solutions.

SEB is the key technological element that allows our LMS and Vocareum to be used safely. Apart from blocking software on students' laptops and non-whitelisted websites, as an added safeguard, SEB provides a `browser exam key' feature that can be used to ensure that submission platforms will only work when accessed \emph{within} SEB. For each exam, a unique (secret) hash key can be generated and provided to external platforms so that they can verify that students are indeed using the lockdown browser. While Vocareum implemented this feature for us, Blackboard Learn does not yet support it (although it does so for the Respondus LockDown Browser~\cite{Respondus}). This meant that we had to develop additional policies (which we discuss later) in order to ensure that the LMS part of the exam is not accessible outside of the exam venue.

\begin{figure}[!t]
	\centering
	\includegraphics[width=\linewidth]{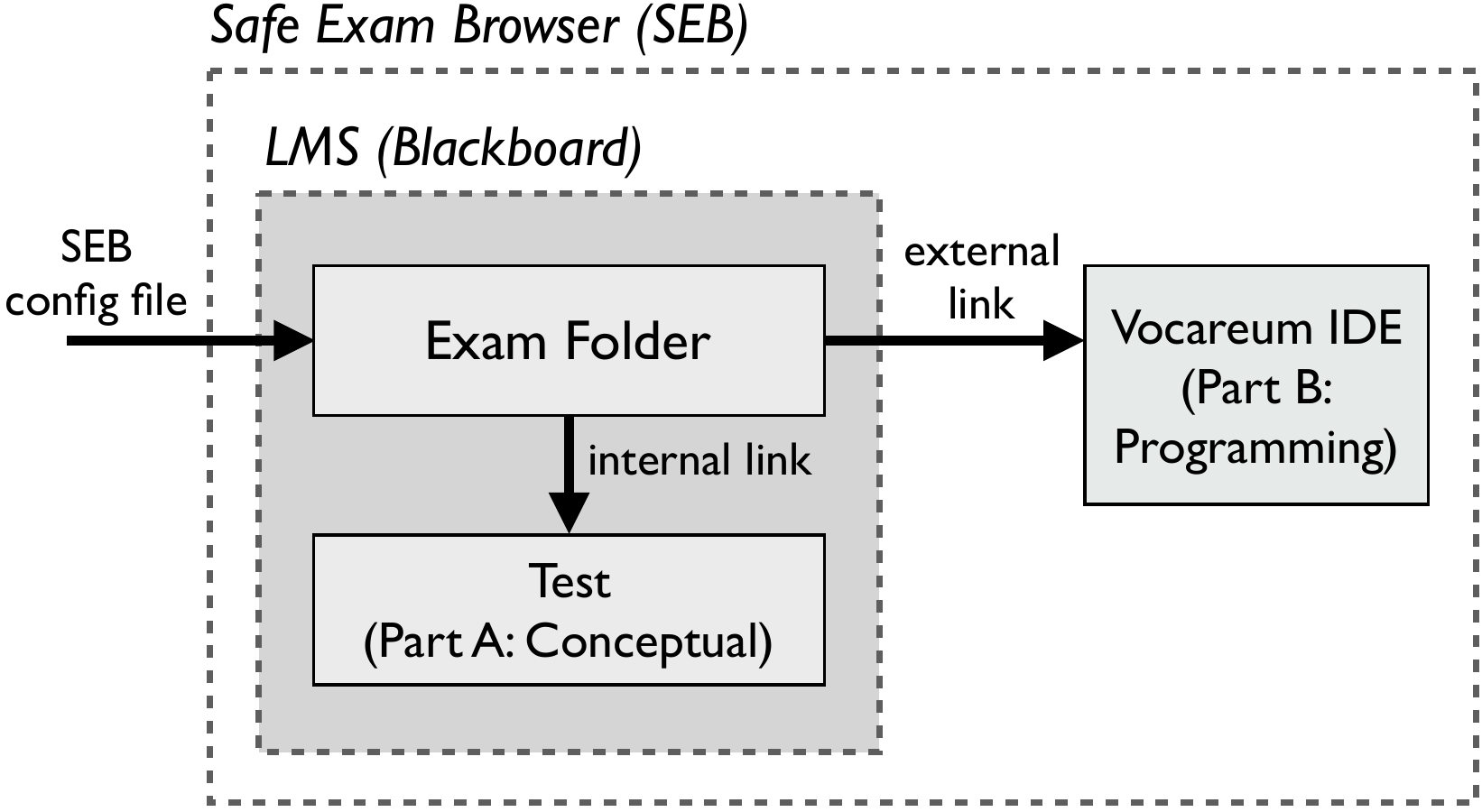}
	\caption{Overview of the technology in our BYOD exam}
	\label{fig:schematics}
\end{figure}

\begin{figure}[!t]
	\centering
	\includegraphics[width=0.9\linewidth]{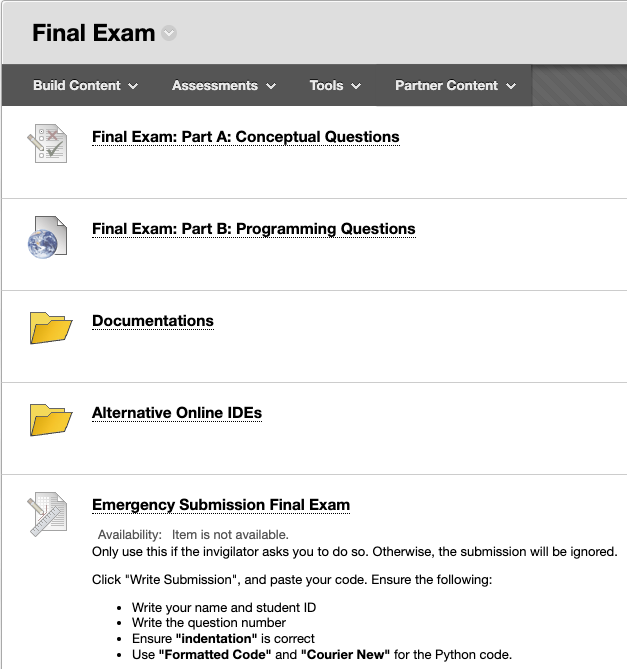}
	\caption{Contents of the Exam Folder in our LMS}
	\label{fig:lms}
\end{figure}

Figure \ref{fig:schematics} provides an overview of how students interact with these technologies during an exam. First, students launch SEB by opening a `config file' specific to the exam that is provided by an instructor. This is configured to launch the homepage of our LMS. After logging in, students can access an Exam Folder (Figure~\ref{fig:lms}) from about 15 minutes before the exam up until about 15 minutes after the end. Once the exam starts, they can click an internal link to launch Part A, consisting of conceptual questions in Blackboard's native `Test'. They can also click an external link to launch the Vocareum platform for attempting and submitting Part B on practical programming. Students can switch between these two parts at their convenience, or even return to the Exam Folder to access additional documentation and links to alternative web-based IDEs.

\begin{figure}[!t]
	\centering
	\includegraphics[width=1\linewidth]{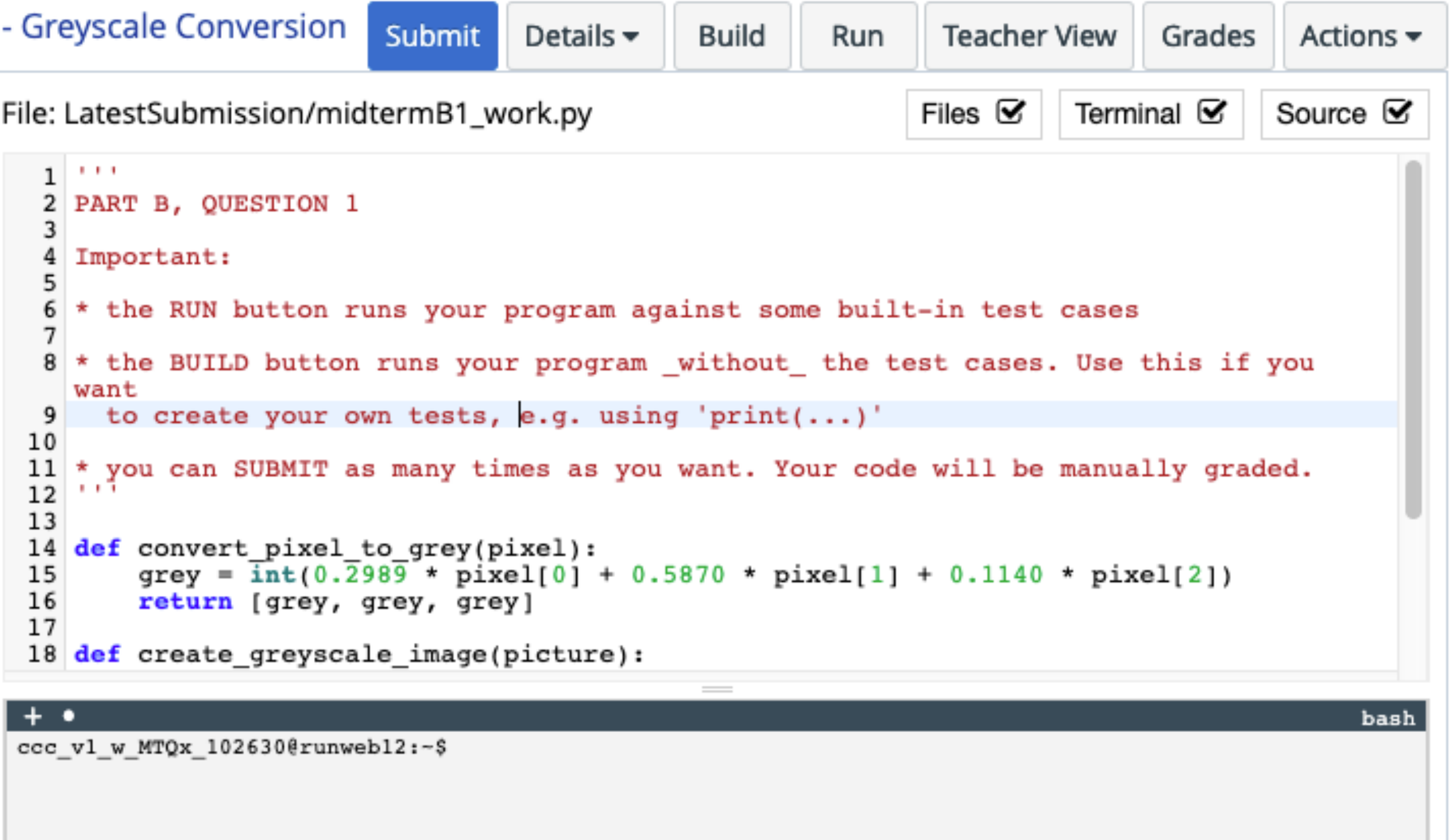}
	\caption{Vocareum web-based IDE and submission platform}
	\label{fig:vocareum}
\end{figure}

In the Vocareum IDE (Figure~\ref{fig:vocareum}), we provide students with some `starter code' (i.e.~a template). Students can test their code by clicking the \textsc{Run} button, which executes all of the built-in test cases provided by the instructors, displaying their output in the terminal. Vocareum also provides a \textsc{Build} button, but since Python does not require compilation, we re-purpose the button to simply run the code without built-in test cases, i.e.~allowing students to write their own tests without the instructors' tests cluttering up the terminal (students can also write tests directly in the terminal). We inform students beforehand on the use of this button and put instructive comments into the starter code (Figure~\ref{fig:vocareum}). Once satisfied, students submit their solutions using the \textsc{Submit} button. In our exams, we configure Vocareum to allow an unlimited number of submissions, and grade only the latest submission that is received. To encourage timely submissions, strict penalties are imposed for students who submit after the exam: 50\% if within 5 minutes, and 100\% thereafter.

After the exam, grading is conducted within the LMS and Vocareum, with the latter allowing instructors to run the students' submissions and quickly assess their correctness. Parts of the grading can be automated (e.g.~based on test cases), but all submissions are manually checked before the final grade is confirmed. We also still run a plagiarism detection tool in Vocareum to mitigate the risk of any potential flaws in the invigilation of the exam or SEB~\cite{Sogaard16a}.

\substepseparator

\noindent\textbf{Policies.} To mitigate the main technological limitation---that Blackboard Learn does not yet support SEB's browser keys---as well as other cheating possibilities, we devised and put in place the following policies as safeguards. First, students are told to arrive at the exam venue early and launch the SEB config file. At this stage, students must wait to enter a \emph{settings password} to activate it, which is provided only 15 minutes before the start of the exam. The questions on our LMS and Vocareum are protected by a separate \emph{test password}, which is provided to students only after the invigilators have verified that all students have launched SEB. The test password is changed 15 minutes after the exam begins and is then known only to the invigilators, preventing students who leave the exam venue (e.g.~for the bathroom) from sharing it with other people, and thus preventing the exam from being accessed remotely.

\begin{figure*}[!t]
	\centering
	\includegraphics[width=0.9\linewidth]{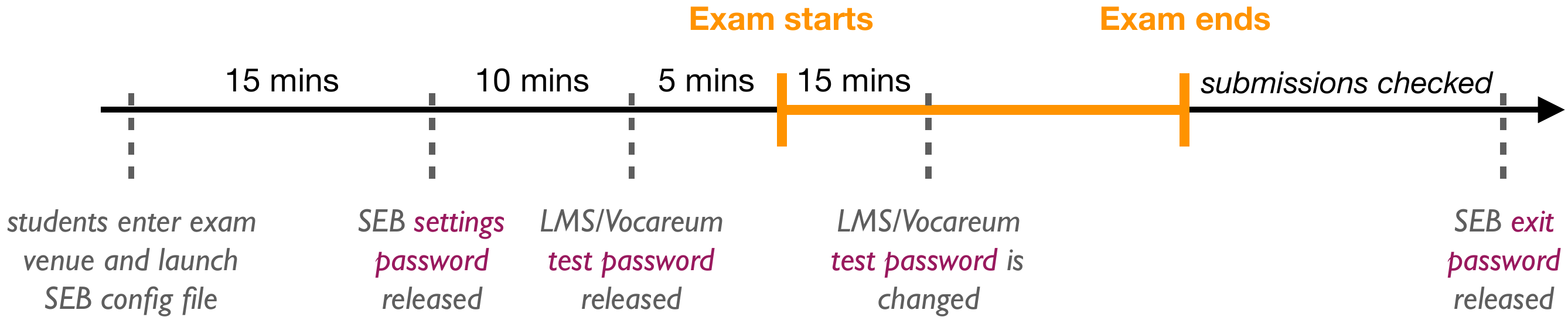}
	\caption{Timeline (not proportional) of the key events in our BYOD programming exams}
	\label{fig:exam-timeline}
\end{figure*}

\begin{table*}[t]
\caption{Examples of policy documents for invigilators to follow} \label{tbl:policy}
\small\begin{tabular}{l|l|l|l}
\textbf{Policy Name}                              & \textbf{Severity} & \textbf{Systems Affected} & \textbf{Policy Contains}                                                                                                   \\
\hline
Normal sequence of invigilation     & ---   &        ---             & venue, timing, contact person, what to do at what time (Fig.~\ref{fig:exam-timeline})                                                     \\
Student forgets their LMS password        & Low      & LMS                 & steps to recover password, rules on time extension                                                            \\
Student forgets their Vocareum password   & Low      & Vocareum            & steps to recover password, rules on time extension                                                        \\
Students cannot run SEB & Medium & SEB, LMS, Vocareum     & steps for backup laptop or alternative exam mode (paper) \\
Vocareum is very slow               & High     & Vocareum            & steps to use alternative online IDE, submission using LMS \\
LMS is not accessible          & High & LMS, Vocareum       & steps for alternative exam mode (paper), contact tech support \\
Total network breakdown                   & High     & LMS, Vocareum       & steps to exit SEB and use alternative exam mode (paper)
\end{tabular}
\end{table*}

To help eliminate the possibility that students access Blackboard Learn questions outside of SEB, we make use of the browser's \emph{exit password} feature. The exit password is required to close SEB, and is communicated only after the invigilators have verified that all solutions have been submitted. Note that in situations where a laptop is forced to restart, SEB will automatically restart too. Even if it fails to do so, students would still need the (changed) test password to access the LMS and Vocareum, thus alerting invigilators to the case. Figure~\ref{fig:exam-timeline} summarises the overall process of running the exam.

Given the different possible points of failure (SEB, LMS, Vocareum, WiFi, students' laptops), we systematically designed procedures for handling all combinations of failures. These involve backup submission platforms, backup IDEs, backup laptops, and in the case of total network failure, manual submission of answers on paper and USB sticks. Some examples of these different cases are summarised in Table \ref{tbl:policy}.

\substepseparator

\noindent\textbf{People.} Finally, it is critical to consider the people involved in running our BYOD solution, especially when exams are run simultaneously across multiple classrooms (as in Digital World). To ensure a consistent experience, invigilators are briefed together in a training session, with special attention given to procedures such as those in Table~\ref{tbl:policy}. Furthermore, our institution's Educational Technology team strongly supports us by conducting mock exams, testing the software, developing/updating policies, and by providing technical assistance to invigilators and students during exams. Finally, we are also fortunate that Vocareum is willing to provide real-time support during our exams to ensure that technical issues related to their platform can be solved immediately.

\section{Reflections}\label{sec:reflections}

In this section, we reflect on how effective our BYOD format was for the 2019 iteration of Digital World. We report on its advantages and disadvantages from three different perspectives. First, our own, as co-designers of the BYOD exam solution; second, those of our co-instructors who helped to run the exams; and third, those of our students, who were ultimately the ones subjected to it. Finally, we present our overall judgement as to whether the solution achieved our goals, and whether its benefits outweighed its drawbacks.

\substepseparator

\noindent\textbf{Our Reflections.} Our first use of the BYOD solution was in a one-hour quiz taken by the entire cohort of students a few weeks before the mid-term. Despite conducting a mock test one week before this quiz, we ran into some difficulties that led to several improvements for subsequent exams. First, with several hundred students using the platform at the same time, Vocareum was not able to cope with the computational load, leading to several students being unable to submit or test their answers. We worked with Vocareum to address the root cause of the issue, and took steps to reassure students of this. We also devised new policies to mitigate any future technical issues with Vocareum, in particular, using Blackboard's native `Assignment' and `Journal' features as a backup means of submission. There were a small number of students who were not able to launch SEB at all during the quiz. They were provided with backup laptops during the test, and our Educational Technology department investigated afterwards. The most frequent reason for the problem was that the wrong version of SEB had been installed.

We found that, despite organising a mock test, several students still struggled with SEB in the first quiz. To resolve this, we started providing `dummy' SEB config files before the subsequent exams, which students could use to test the interface and practice programming within its restrictions. We observed that this improvement increased the students' confidence and familiarity with the exam environment. Related to this issue, the University of Tasmania (UTAS) reports that for their Linux USB BYOD exams, students are required to test the exam environment beforehand at an organised workshop or on their own~\cite{UTAS}. To take part in the exam, students must prove that they completed the testing by producing a certificate that is issued upon its completion. In future iterations of our course, we will consider whether or not to introduce a similar policy to minimise disruption caused by unfamiliarity.

In subsequent exams and quizzes, our BYOD solution appeared to work smoothly, aside from a few isolated cases which were managed by invigilators and Educational Technology staff according to our policies (an example of how the components of Figure~\ref{fig:aspects} are interdependent). One additional issue that arose, however, was related to the security of websites and tools launched by SEB. While SEB provided secure access to Vocareum, it did not prevent Vocareum from circumventing the restrictions. We discovered, for example, that it was possible for students to use the Vocareum terminal (see the bottom of Figure~\ref{fig:vocareum}) to make external connections. Fortunately, Vocareum responded quickly by providing an option to run our assessments on an `Exam Server' that did not allow any internet access. Our lesson here is to always challenge our assumptions about the external tools enabled for use within SEB, and to work with vendors to make sure that they are properly secured. An alternative solution would have been to provide offline IDEs via SEB and a virtual desktop infrastructure (as in some lab-based ETH~Z\"{u}rich exams~\cite{Halbherr-et_al14a,Luethi-et_al19a}), where we would have full control, but also the time-consuming task of ensuring that the virtual machines are properly secured.

Hillier and Fluck \cite{Hillier-Fluck13a} mentioned clear user guides as a key requirement for BYOD exam formats. For their USB BYOD exam, the guides written for the various stakeholders are provided online \cite{TransformingExams}. In our course, we prepared slides to be shown during the exam containing the steps and passwords. Moreover, after our first quiz, we also communicated to students our procedures and written policies in case of failure. We believe that together, these documents provided to students meet the requirement of clear user guides. In future iterations of our course, our documentation will continue to be revised for better clarity, and we will aim to communicate our procedures to students more clearly in advance of the exams.


\noindent\textbf{Instructor Reflections.} We elicited a mix of positive and negative views from the instructor team, including a more critical viewpoint (paraphrased): \emph{``Just to handle some dishonest students, we are inconveniencing the majority of honest students in their exam, and I think this is no good. Why do we have to give pain to the majority of honest students just for some dishonest ones?''}

While agreeing that it was good for students to be able to complete their programming exams on their own devices, this faculty member felt that our BYOD format involved too much inconvenience, and prevented students from using their day-to-day offline IDEs (e.g.~Spyder). The majority of other instructors, however, felt that it was important to implement measures to prevent cheating, in order to reassure the honest students that the exam was being conducted fairly. It was also argued that eliminating the possibility to cheat helps remove the temptation to do it in the first place, especially for students who are struggling under heavy workloads. Still, the discussion led us to reflect that we should always seek to minimise the disruption to students in setting up these safeguards. Allowing students to practice with SEB before the exam is one such way for us to achieve this.

A feature of the exam format that instructors particular liked (over, e.g.~pen and paper exams) was the fact that Vocareum allowed them to complete grading faster by having the code ready to run, along with the results of some built-in test cases to indicate the likely quality of code. Note, however, that instructors manually inspected all submissions, to ensure that students did not simply hardcode return values for the given test cases. (Some always try!)

\substepseparator

\noindent\textbf{Student Reflections.} To gain some viewpoints of the students, we conducted an informal online survey with 13 of them who were reachable over the summer vacation. Using open-ended questions, the students were asked about the positive and negative aspects of the exams, whether they preferred a BYOD or pen and paper format, and if they had any suggestions for improvement.

In terms of the positive aspects, there was a very clear message from the survey responses:

\begin{center}
\noindent\fbox{%
    \parbox{0.76\linewidth}{%
        \small\emph{The students were positive about being able to use their own laptops, as they were familiar and comfortable with them.}
    }
}
\end{center}

\noindent Some also mentioned that they can type faster on their own laptops, and need not worry about getting used to the idiosyncrasies of other computers. All of them preferred BYOD exams, because it is possible for them to test and debug their code in the same way as practiced during regular lessons. A few responses highlighted that debugging is an important skill and that paper exams would not allow such a skill to be tested. Several remarks also mentioned that BYOD exams were more realistic, with one student writing: \emph{``nobody outside actually programs on pen and paper''}. 

Some negative responses tended to focus on Vocareum, with some students remarking on its slow performance during the first quiz. Fortunately, we did not see any major complaints about the other assessments, as Vocareum were able to rectify the issue for subsequent exams. Moreover, Vocareum is simply an external component of our BYOD exam and any issues can be addressed independently: they are not related to the use of lockdown browsers. Two remarks addressed the user experience of SEB, mentioning that it is difficult to switch between tabs. This affirms that it is necessary to allow students time to familiarise themselves with SEB on their own. A minority of remarks cited the preparation steps, such as installing the lockdown browser and downloading new config files for every exam, as a negative.

Surprisingly, none of the students in the survey complained about the exam procedures themselves (such as the use of several passwords), and no suggestions for improvement were made about them. Some instructors were worried that the procedures and policies might create a negative experience during the exam. A particular concern was the incompatibility of lockdown browsers with password managers, which are often used by our students. This incompatibility meant students had to memorise or write down their personal LMS and Vocareum credentials (often long and random) for use in the exam. This problem is made worse by an IT policy at our institution that locks students' university accounts after three failed login attempts---a possible source of anxiety for examinees. However, from the students that were surveyed, none of them mentioned this issue at all. It is possible that it was minimised by our frequent reminders to memorise or write down their passwords, and also by the provided `dummy' SEB config files (in which they would have seen the issue first-hand). Furthermore, by requiring students to come 30 minutes early, our trained invigilators were able to solve isolated cases before the exam started.

\substepseparator

\noindent\textbf{Overall Judgement.} Taking all viewpoints into account, we felt that the effort required to design and implement this BYOD solution was worth the results it provided us with. First, as instructors it provides us with confidence in the integrity of all submissions received. Second, for students, it gives them the convenience of using their own devices and familiar programming environments (all while remaining controlled). Finally, the convenience of being able to use technology in the grading process (fully automated in the LMS; partially automated in Vocareum) allows for instructors to optimise their time and for students to receive feedback faster.

\section{Conclusion}\label{sec:conclusion}

In this article, we have described a BYOD exam solution for conceptual and practical programming questions, combining technologies such as lockdown browsers, LMSs, and web-based IDEs. We reflected on our implementation of this format in a major undergraduate programming course, finding that students were positive about being able to use their own devices, and were accepting of the exam procedures necessary to facilitate this. We learnt that technology is just one of the key components necessary for a secure BYOD programming exam, and that developing proper policies, safeguards, and training is equally important. With all these considerations in place, our experience has shown us that (despite some initial difficulties) our BYOD exam format is a viable compromise solution that balances our aim for pedagogically-aligned assessments with the need to prevent cheating and plagiarism.

\begin{acks}
	First, we would like to thank the Educational Technology team at SUTD (especially Magnus Bengtsson and Kristy Chan Hui Ling) for their consistent and strong support in helping to develop this BYOD solution. Second, we would like to thank the instructors and students of Digital World for their patience, support, and feedback as we implemented these changes to the exams. We would especially like to thank those students who responded to our informal survey during their break. Finally, we would like to express our gratitude to the anonymous SIGCSE referees for their helpful suggestions.
\end{acks}

\bibliographystyle{ACM-Reference-Format}
\balance
\bibliography{references}

\end{document}